\documentclass[12pt]{iopart}
\usepackage[dvips]{graphicx}
\usepackage{setstack}

\begin{document}
\title{Critical behaviour of fully-frustrated Potts models}
\author{D P Foster, C G\'erard and I Puha } 
\address{Laboratoire de Physique Th\'eorique et 
Mod\'elisation (CNRS ESA 8089)\\
Universit\'e de Cergy-Pontoise,
5 Mail Gay-Lussac\\ 95035 Cergy-Pontoise Cedex, France}

\begin{abstract}
A numerical transfer matrix calculation is presented for two
fully-frustrated three-state Potts models on the square lattice: the
Potts piled-up-domino model and the Potts zig-zag model. 
The ground state entropies and phase diagrams are found. 
The Potts piled-up-domino model
displays a finite-temperature transition when the frustration effects
are maximal, and displays reentrant behaviour, in contrast to the
Ising model equivalent. The ground-state entropy per spin is 
larger than in the Ising equivalent. 
The Potts zig-zag model displays the same qualitative
behaviour as its Ising counterpart, and the ground-state entropy per
spin is the same in the Potts and Ising cases.
\end{abstract}

\pacs{05.20.-y, 05.50.+g, 64.60.Fr, 75.10.Hk}


\maketitle

\section{Introduction}

Spin models are said to be frustrated when they contain competing
interactions which prevent the simultaneous minimisation of 
all the lattice bonds in the ground state\cite{gt77}. 
The system is additionally described as fully frustrated if each
plaquette of the lattice is itself frustrated\cite{jv85}.
The
nature of a phase transition depends partly on the symmetry
group of the candidate low-temperature phases.
Frustration effects are then of most
interest when they  lead to infinitely degenerate
or ``nearly-degenerate'' ground states.
These will in general satisfy a larger symmetry group than the
individual spins, possibly giving rise to exotic critical phenomena
\cite{dpbook}.  
One such phenomenon is re-entrance\cite{dpbook1}. This occurs when
a disordered paramagnetic phase comes to lie at a lower temperature
than  an ordered
phase; as the temperature is increased there is a transition from
disorder to order. One possible 
explanation for this phenomenon is the so-called ``order by disorder''
mechanism\cite{vbcc80}.

In this article we present results for two fully-frustrated three-state  Potts
models: the Potts Piled-Up-Domino Model (from here on referred to
simply as the piled-up-domino model) and the Potts Zig-Zag Model (or
more concisely the zig-zag model). These are
natural extensions of the Ising-model equivalents studied previously by
Andr\'e {\it et al.}\cite{abcc79} and are defined as follows:
 
The $q$-state Potts model is defined as a set of  variables
$\{\sigma_i\}$ associated with the sites $\{i\}$
of a lattice\cite{potts}. Each $\sigma_i$
takes one of $q$ distinct
values. The Hamiltonian of the model is given by:
\begin{equation}
{\cal H}= -\frac{1}{2} \sum_{i,j} J_{i,j}
\delta_{\sigma_i,\sigma_j},
\end{equation}
where the sum runs over all pairs of sites and $J_{i,j}$ are the
interaction strengths between a given pair of spins. 
For the models of interest in this article, 
the interaction strengths are given by 
\begin{equation}
J_{i,j}=\cases{
0&for
non-nearest-neighbour spins,\\
J_1&for n.n. spins along the solid lines,\\
J_2&for n.n. spins along the dashed lines,
}
\end{equation}
as shown in figures~\ref{lattice}~(A) and~(B) for the
Piled-Up-Domino 
and
Zig-Zag models respectively. 
For convenience we define
$\alpha=J_2/J_1$. $J_1$ will be taken as positive throughout. In this
article we shall concentrate on the $q=3$ Potts models.

In the Ising versions of these models, the frustration effects
suppress the critical temperature, which becomes zero when
$\alpha=-1$\cite{jv77}. 
We show that, while 
this remains the case for the three-state Potts 
zig-zag model, the three-state Potts piled-up-domino model
has a finite
critical temperature when $\alpha=-1$ and displays reentrant behaviour.

The article is divided as follows:
In the next section we present
briefly the transfer matrix method used in our study, and 
in  \sref{gs} we apply
this method to the calculation of the residual ground-state entropy
for the piled-up-domino model. We show that 
the residual ground-state entropy for
the zig-zag model is
exactly related to the equivalent Ising-model case. In
\sref{pd} we present the finite-temperature phase diagrams for
the two cases and \sref{conc} will be devoted to discussion.

\section{The Transfer Matrix Method}

The results presented in this article 
are obtained using a (numerical) transfer matrix calculation, giving
the 
partition function for strips of finite width $L$ but
infinite length, $N\to\infty$. The results are therefore numerically
exact for a given lattice width $L$. 

The partition function for the model may be written in the form
%
\begin{equation}
{\cal Z}=\sum_{\{\sigma\}}\prod_{x} V({\cal C}_x)H({\cal C}_x;{\cal C}_{x+1}),
\end{equation}
%
%
where 
$V({\cal C}_x)$ is
the contribution to the 
Boltzmann weight of a given column spin-state ${\cal C}_x$
due to interactions within a column and $H({\cal C}_x;{\cal C}_{x+1})$ 
is the contribution due to interactions between adjacent
columns, and depends on the spin states of the two columns
${\cal C}_x$ and ${\cal C}_{x+1}$.
If additionally periodic boundary conditions are taken in the
$x$-direction, the partition function for the
system may be written as a trace over a product of transfer matrices:
\begin{equation}
{\cal Z}=\Tr \{{\cal T}^N\},
\end{equation}
where ${\cal T}=V^{1/2}HV^{1/2}$ 
and the sum over spin states is taken care of by the 
 matrix
multiplications.
Denoting the
eigenvalues of ${\cal T }$  by $\lambda_i$,
\begin{equation}
{\cal Z}=\sum_{i} \lambda_{i}^{N}.
\end{equation}

As $N$ becomes large the partition function is dominated by the
largest eigenvalue. The (dimensionless) free energy per spin, $f$ 
is given by
\begin{eqnarray}\nonumber
f&=&\lim_{N\to\infty}\frac{1}{NL}\log({\cal Z}),\\\label{fen}
&=&\frac{1}{L}\log(\lambda_0),
\end{eqnarray}
and the correlation length is given by 
\begin{equation}\label{cor}
\xi=\log\left(\frac{\lambda_0}{|\lambda_1|}\right),
\end{equation}
where $\lambda_0$ is the largest eigenvalue and $\lambda_1$ is the
second largest eigenvalue (in absolute value)\cite{jmy}.

\begin{figure}
\begin{center}
\includegraphics[width=10cm]{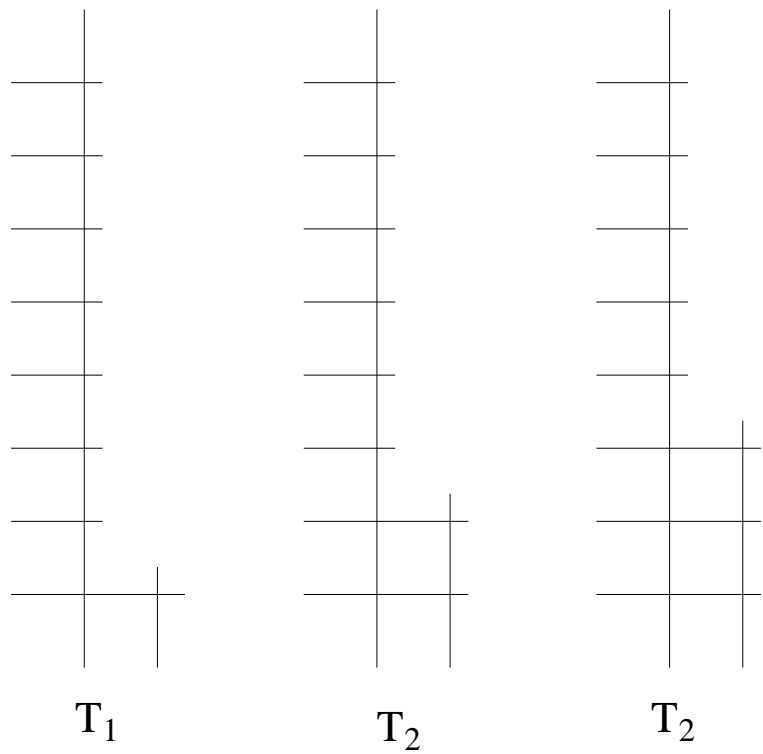}
\end{center}
\caption{Construction of the transfer matrix by single spin
addition}\label{transmat} 
\end{figure}

The transfer matrix ${\cal T}$ is indexed by ${\cal C}_x$ and ${\cal
C}_{x+1}$
both of which correspond to one of $3^{L}$ distinct
column spin states. The capacity required to store the $3^{2L}$ 
matrix elements on a computer soon becomes prohibitive.
In order to extend the range of values of $L$ which may be treated
numerically, the transfer matrix ${\cal T}$ is decomposed into a
product of sparse matrices as follows:

One may view the transfer matrix as an operator which adds an
additional column to the lattice. Each element of the matrix is
indexed by the spin configuration on the last column of the lattice
($L$ spins) and the spin configuration of the column being
added. If we define a dangling bond as the bond which has only one
spin attached to it, we see that the transfer matrix replaces the $L$
existing dangling bonds with $L$ new ones. The entry configurations and
exit configurations are given by the values of the spins attached to
these dangling bonds.
If now, instead of adding a whole column at once, one spin is added at
a time, there are still $L$ dangling bonds beforehand and afterwards
(see \fref{transmat}).
This operation may be written as a matrix indexed by the configuration
of the $L$ spins
connected to a dangling bond before the addition of the new spin, and
the configuration of the $L$ spins connected to a dangling bond 
after the spin
addition. Since $L-1$ spins are common to the input and output
configurations, the matrix elements will be non-zero only 
if these spins have the same values in the input
and output configurations. The corresponding matrix is thus sparse.
In
this way
 ${\cal T}$ may be
decomposed: ${\cal T}={\cal
T}_1{\cal T}_2^{L-2}{\cal T}_3$ where the matrices 
${\cal T}_1$, ${\cal T}_2$, ${\cal
T}_3$ correspond to one-spin additions. 
If the boundary conditions and in-column interactions are taken care
of when the first and last spins are added, then the intermediary spin
additions may all be performed by the same matrix.
The combination of the ${\cal T}_i$ being sparse and the small number
of distinct matrices required to build up ${\cal T}$  
vastly reduces the storage requirements 
and enables calculation for lattice widths up to $L=12$.

An alternative method for writing the transfer matrix involves the
Kasteleyn-Fortuin mapping~\cite{kf69} 
to map the Potts model to a lattice animal
problem~\cite{blote}. 
This mapping requires periodic boundary conditions to enable
a sparse-matrix reduction~\cite{blote}. 
For $q=3$ the transfer matrices in the two representations
require an
equivalent amount of storage but in a spin representation it is easier
to implement the different boundary conditions used in this article.

The eigenvalues used were calculated using 
the Lanczos method, using the ARPACK package. 

\section{Ground-State Properties}\label{gs}

In this section we calculate the ground-state entropy for the two
models.
In both models, when $\alpha>0$ the model is ferromagnetic. 
While they are anisotropic,
there is 
no competition in the interactions. The system minimises its energy by 
choosing the same state for all the spins giving $q$ distinct
ground states\cite{potts,wu}. The entropy per spin is zero.

A lattice bond is ``satisfied'' when the interaction energy
along it is minimal, and ``frustrated'' otherwise.
It may be seen that for $\alpha<0$ it is no longer
possible to simultaneously satisfy all
four bonds of a plaquette. The model is then said
to be 
frustrated. However, if 
$\alpha>-1$ there is no ambiguity, in the ground state,
over which bond to frustrate. The 
frustration does not change the ground state configuration, and 
the entropy
per spin remains zero.

For $\alpha=-1$ the energy of a plaquette is minimal 
regardless of which
bond is frustrated as long as one and only one
bond is frustrated.
There are an infinite number of  
frustrated bond configurations satisfying the ground state
condition and the entropy per spin in this case is finite.

For $\alpha<-1$ the ground state degeneracy is still infinite. While 
it is clear
that the system will satisfy the antiferromagnetic bond in order to 
minimise the energy, there is still a choice over which of three  
ferromagnetic bonds to frustrate. 
The entropy per spin here is also finite.

\subsection{Piled-up-domino model}

The ground-state entropy for the piled-up-domino model was calculated
using transfer matrices. 
In the ground state 
it is necessary 
to add
two spins at a time in order to ensure
that only one bond per plaquette is frustrated. Since the energy is
constant and minimal, the entropy per spin is simply the number of
ground states per spin, leading to:
\begin{equation}
s=\frac{1}{L}\log(\lambda_0).
\end{equation}
The entropy was calculated with transfer direction parallel to the
dashed lines as shown in \fref{lattice}. Both free and periodic 
boundary conditions
were considered.
The results are shown in \tref{pudent}.
It may be seen that the 
entropy per spin is substantially larger than for the
Ising-piled-up-domino model.

The extrapolations to $L\to\infty$ throughout the article are
done using the Bulirch and
Stoer algorithm\cite{bst}.

\subsection{The zig-zag model}

\begin{figure}
\begin{center}
\includegraphics[width=12cm]{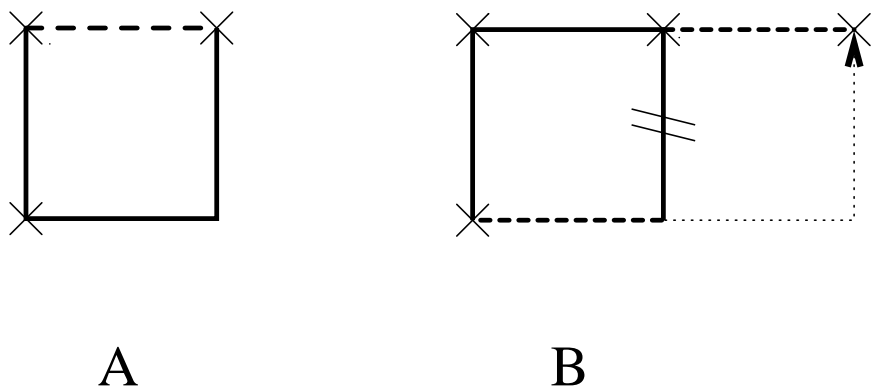}
\end{center}
\caption{Plaquettes for the construction of the 
zig-zag model. Solid lines show interactions of strength $J_1$ and
dashed lines interactions of strengh $J_2$.
The arrow shows path to prefixed spin in the case that
the marked bond is frustrated.}\label{zigplaque}
\end{figure}

The ground-state entropy per spin for the zig-zag model is the same in
the Potts and Ising cases. This can be illustrated
as follows: Imagine that, from some suitable boundary configuration of
spins, a ground-state configuration of spins is built up by adding
alternately plaquettes of type {\bf A} and {\bf B}, shown in
\fref{zigplaque}, in rows, starting from top left and ending
bottom right, one
plaquette at a time. The spins marked by a cross are shared with
plaquettes already added to the lattice, and are thus already fixed. 
In the ground state only one bond per plaquette is
frustrated. Whichever 
of the bonds is frustrated in a plaquette of
type {\bf A}, the ``free'' spin is connected by a ferromagnetic path
to a predetermined spin. At first sight this is not true for plaquettes
of type {\bf B}, however the frustrated bond is shared 
between two plaquettes, and it can be seen that the
``free'' spin  is connected by a ferromagnetic path 
to a predetermined spin either directly,
or via the
neighbouring plaquette (of type {\bf A}) see \fref{zigplaque}.
The entropy per spin in 
the ground
state is thus determined uniquely by the configurational entropy of
the frustrated bonds, with no additional (local)
degeneracy due to the
number of Potts states. This was not so in the piled-up-domino
case where it is possible to create free domains of spins. The
residual ground-state entropy per spin is thus the same as reported for
the Ising model\cite{abcc79}, i.e.:

\begin{eqnarray*}
s&=&\frac{1}{4\pi^2}\int_0^{\pi}{\rm d}\theta\int_0^{\pi}{\rm d}\phi
\log\left[1+4\cos\theta\cos\phi+4\cos^2\theta\right]\\
&=&0.1615\cdots\ {\rm for\ } \alpha<-1\\
s&=&\frac{1}{4\pi^2}\int_0^{\pi}{\rm d}\theta\int_0^{\pi}{\rm d}\phi
\log\left[4(\cos^2\theta+\sin^2\phi)\right]\\
&=&\frac{G}{\pi}=0.2916\cdots\ \ {\rm for\ } \alpha=-1
\end{eqnarray*}

The difference between the ground-states of the Ising and Potts
versions of the zig-zag models is an overall degeneracy relating to
the symmetry of the spin ($Z_2$ and $Z_3$).
This was not the case for the piled-up-domino model, where the entropy
per spin for the Potts case was significantly larger than the for
Ising case for both $\alpha=-1$ and $\alpha<-1$. 
This difference indicates that
the difference in symmetry group for the ground-states in the two
cases is not simply due to  
the difference in the symmetry groups of the spins.
Based on these results, it is reasonable to expect that the phase
diagram of the Potts zig-zag model should be qualitatively similar to
the Ising case, but that the finite-temperature behaviour
of the piled-up-domino model may well be qualitatively different.
This is indeed what will be observed in the next section.

\section{The Finite-Temperature Phase diagrams}\label{pd}

In this section we present the phase diagrams for the piled-up-domino
and zig-zag models.
The critical lines may be identified using
phenomenological renormalisation\cite{mpn76}. At a critical
point the correlation lengths for two strip widths, measured as a
fraction of the strip width, must be the same. This reflects the scale
invariance of a critical system. In practice a finite width system is
always off-critical, however
\begin{equation}\label{fss} 
\frac{\xi_L(T^*)}{L}=\frac{\xi_{L^{\prime}}(T^*)}{L^{\prime}}
\end{equation}
gives a finite-size estimate of the critical temperature, $T^*$, which will
tend to the true critical temperature as the strip widths tend to
infinity.
Assuming that the system is sufficiently close to the true critical
point, the correlation length behaves (to leading order) as
\begin{equation}
\xi_L=A|T^*-Tc|^{-\nu}.
\end{equation}
Since it is the finite width of the system which prevents the
correlation length from diverging, at $T^*$ we have 
$L\propto\xi_L$. If these
two scaling laws are admitted, then a finite-size estimate of the
correlation length exponent may be calculated from the equation:
\begin{equation}
\frac{1}{\nu_{L,L^{\prime}}}=\frac{\log\left(\frac{{\rm d}\xi_L}{{\rm
d} T}/\frac{{\rm d}\xi_{L^{\prime}}}{{\rm
d} T}\right)}{\log\left(\frac{L}{L^{\prime}}\right)}-1.
\end{equation}

\subsection{The zig-zag model}

\begin{figure}
\begin{center}
\includegraphics[width=10cm,angle=-90]{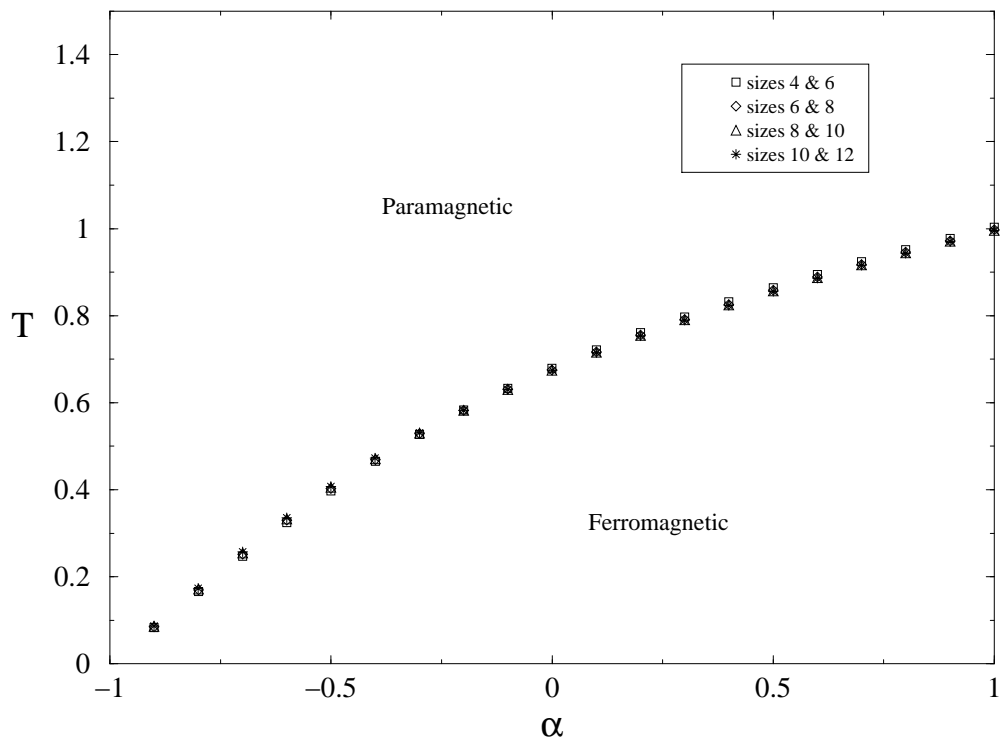}
\end{center}
\caption{Phase diagram for the Potts Zig-Zag model}\label{zigpd}
\end{figure}

The phase diagram found using phenomenological renormalisation for the
zig-zag model with periodic boundary conditions is
shown in \fref{zigpd}. 
The finite-size effects are very small.
The critical temperature is known exactly for two points on the phase
diagram\cite{wu,rgb70}: 
$\alpha=1$, corresponding to the pure ferromagnetic Potts model
on the square lattice, and $\alpha=0$, corresponding to the ferromagnetic
Potts model on the hexagonal lattice. Our numerical values calculated
at these two points are
extrapolated and compared with the exact values in \tref{zigtc}.

Qualitatively the phase diagram is the same as for the Ising
model: There is a phase transition between a low temperature
ferro-magnetic phase and a high temperature phase for $\alpha>-1$. The
critical temperature becomes zero at $\alpha=-1$ and there is no
evidence of a phase transition for $\alpha<-1$. 

\begin{figure}
\begin{center}
\includegraphics[width=10cm,angle=-90]{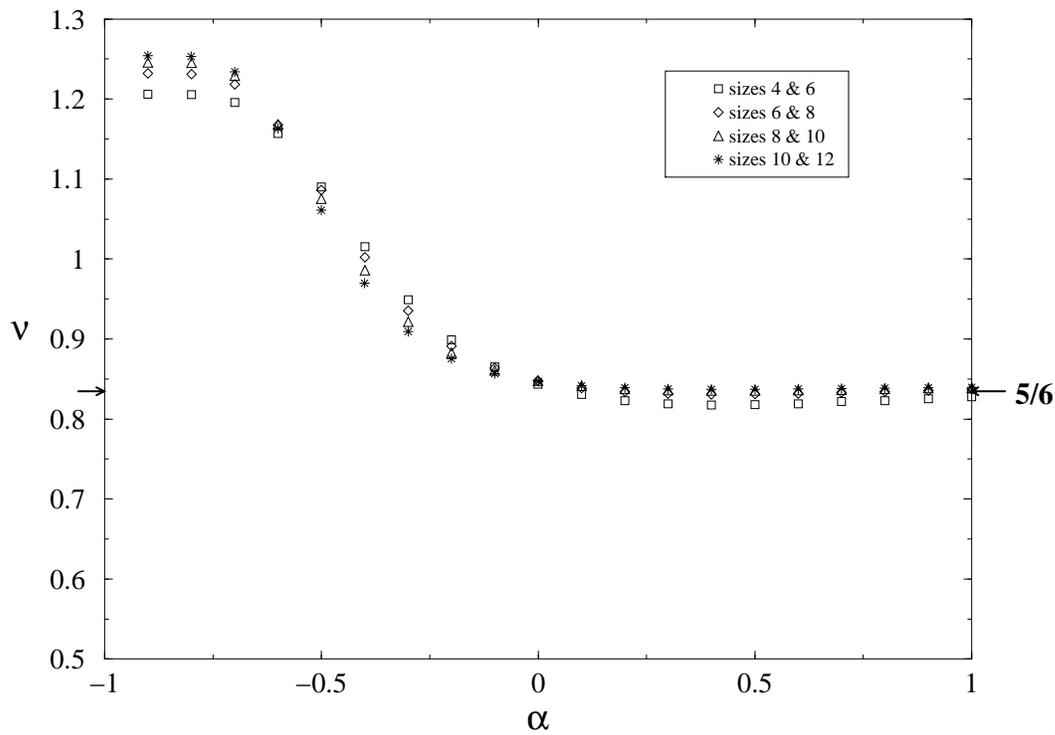}
\end{center}
\caption{Finite size estimates for the exponent 
$\nu$ for the Potts zig-zag model}\label{zignu}
\end{figure}

The
finite size estimates for the exponent $\nu$ are shown in
\fref{zignu}. 
For a pure three-state Potts model ($\alpha=1$
corresponding to the square lattice, and $\alpha=0$ corresponding to the
hexagonal lattice), 
the
value of $\nu=5/6$ is known exactly\cite{mpmdn79}. 
For $\alpha\geq 0$ the estimates for $\nu$
converge to the exact value. Close to $\alpha=-1$ the estimates for
$\nu$ seem to take on a different value, but the range of values of
$\alpha$  over
which this variation is observed diminishes as the system size
increases. There is no evidence of a multicritical point in the phase
diagram between $\alpha=1$ and $\alpha=-1$, and so it would seem
reasonable to
assume that the whole line up to $\alpha=-1$ 
remains in the same universality class as
the ferromagnetic three-state Potts model.

\subsection{The piled-up-domino model}

\begin{figure}
{\large (A)}
\begin{center}
\includegraphics[width=10cm,angle=-90]{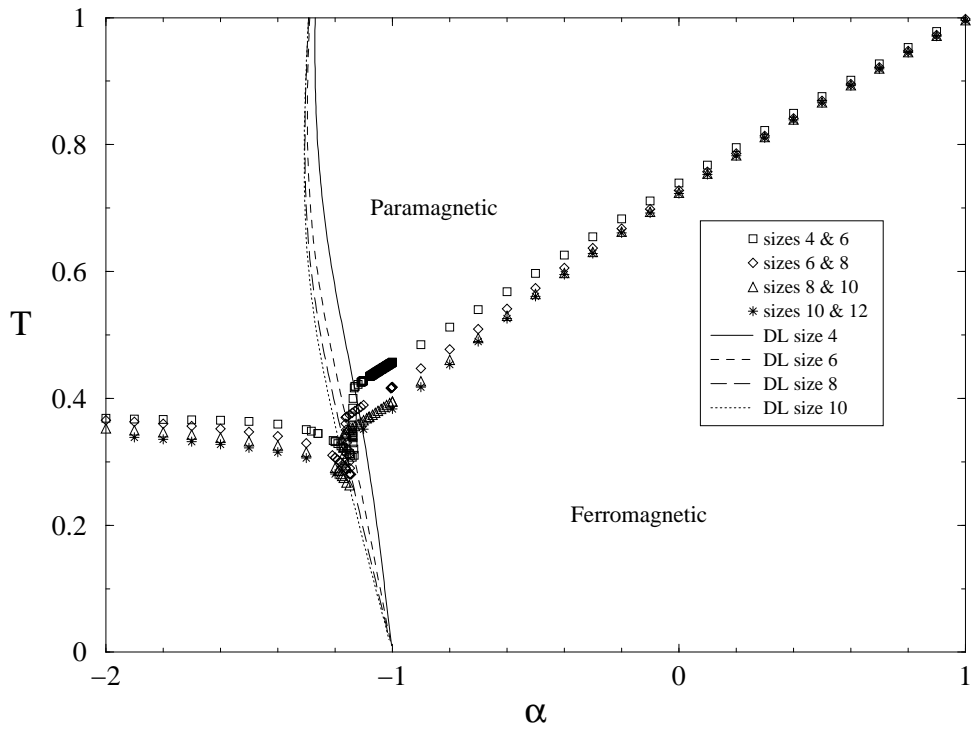}\\
\end{center}
{\large (B)}
\begin{center}
\includegraphics[width=10cm,angle=-90]{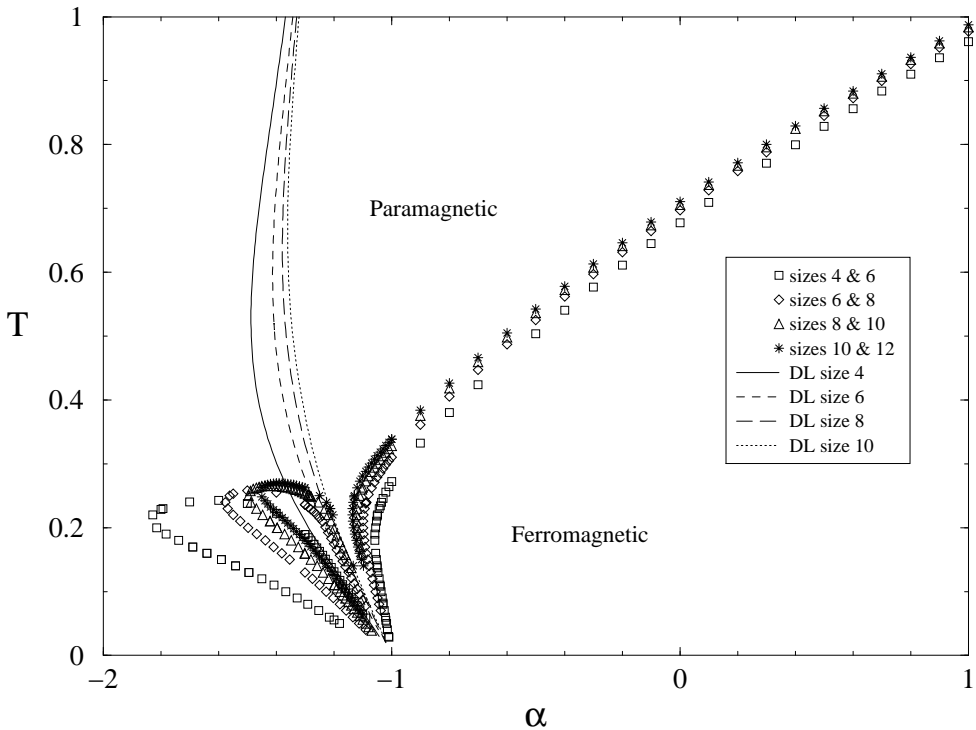}
\end{center}
\caption{Phase diagram for the Potts piled-up-domino model with: (A) periodic 
boundary conditions and (B) free boundary conditions. The disorder
lines are shown as lines, and the phase boundaries as symbols.
The numerical uncertainty is smaller than the size of the symbols.}
\label{pudpd}
\end{figure}

The phase diagrams for the piled-up-domino model with periodic
and free boundary conditions 
found using
phenomenological renormalisation are shown in \fref{pudpd}.

\begin{figure}
\large{(A)}
\begin{center}
\includegraphics[width=10cm]{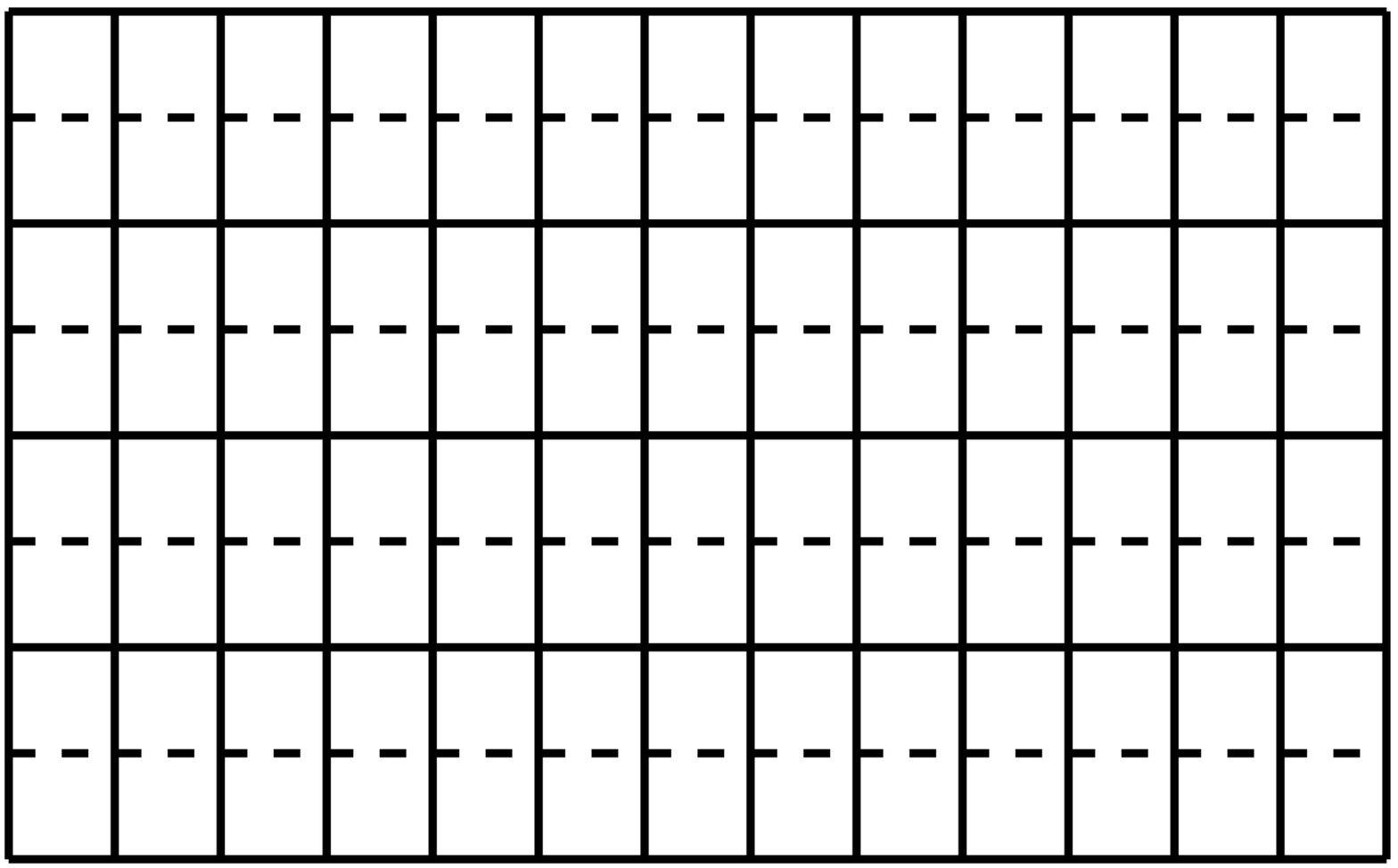}\\
\end{center}
\large{(B)}
\begin{center}
\includegraphics[width=10cm]{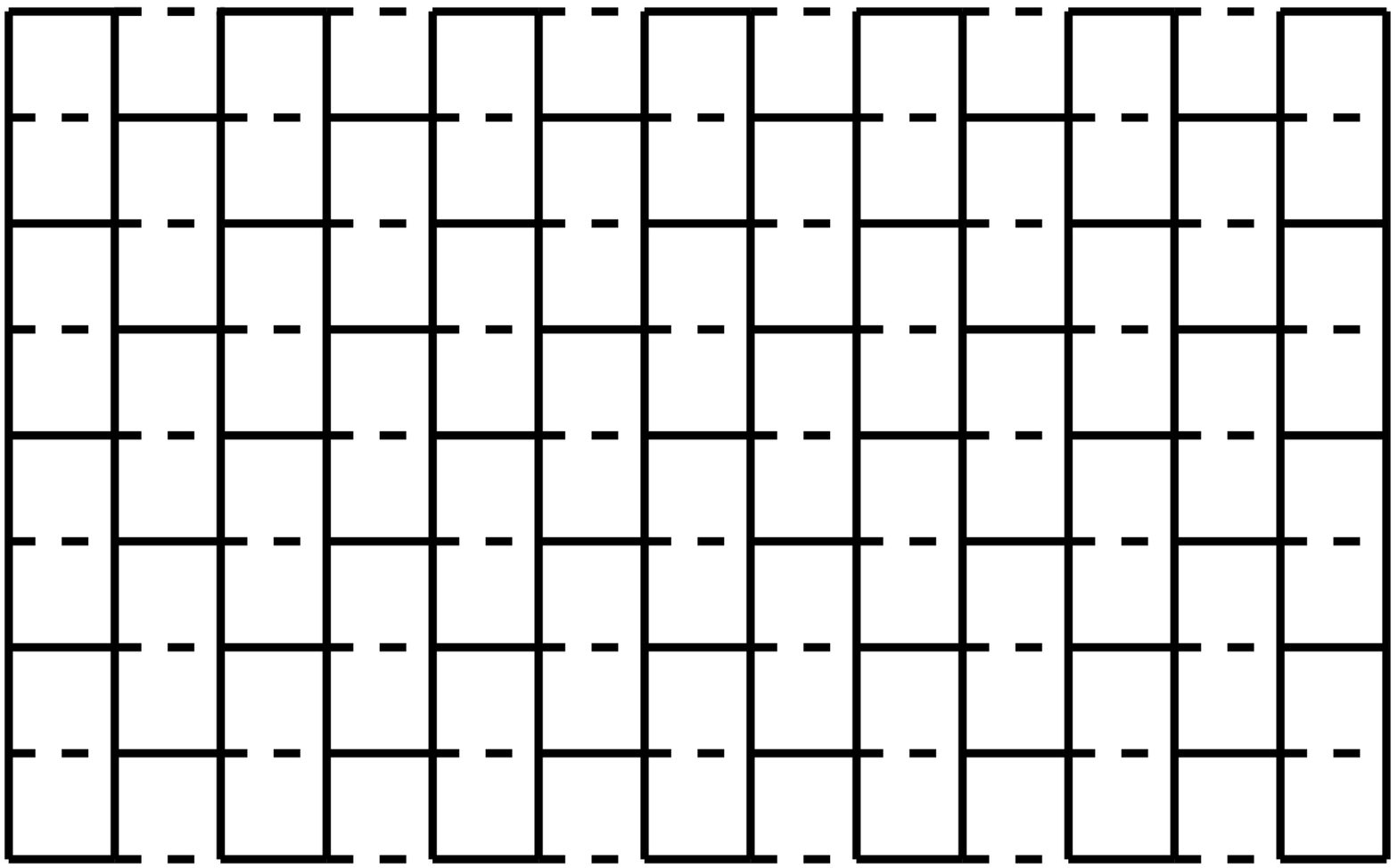}
\end{center}
\caption{The piled-up-domino model (A) and the zig-zag model (B). 
The interaction energy is $J_1$
along the solid lines and $J_2$ along the dashed lines.}\label{lattice}
\end{figure}

The critical temperature is known
exactly for $\alpha=1$ \cite{wu}, 
the pure three-state Potts model on the square
lattice, and at this point the critical temperature coincides trivially
with that found in the zig-zag model.
It may also be easily determined for $\alpha=0$.
When $\alpha=0$
the spins on the dashed lines (see \fref{lattice}~(A)) may be
traced over. The resulting model is a ferromagnetic Potts model on a
square lattice with
different interactions in the  $x$ and $y$ directions: $J_x=J_1$ and 
\[
\exp(\beta J_y)=\frac{\exp(2\beta J_1)+2}{2\exp(\beta J_1)+1}.
\]
The critical temperature is known to satisfy\cite{wu}:
\[
\left(\exp\left(\beta J_x\right)-1\right)\left(\exp\left(\beta J_y\right)-1\right)=3,
\]
which leads to
$T_c=1/\log(4)$, measured in units of $J_1/k$. In \tref{pudtc} we
compare the finite-size critical-temperature estimates, extrapolated
using the 
Bulirch and
Stoer algorithm\cite{bst}, with the exact value calculated at
$\alpha=0$. The agreement is good.

\begin{figure}
{\large (A)}
\begin{center}
\includegraphics[width=10cm,angle=-90]{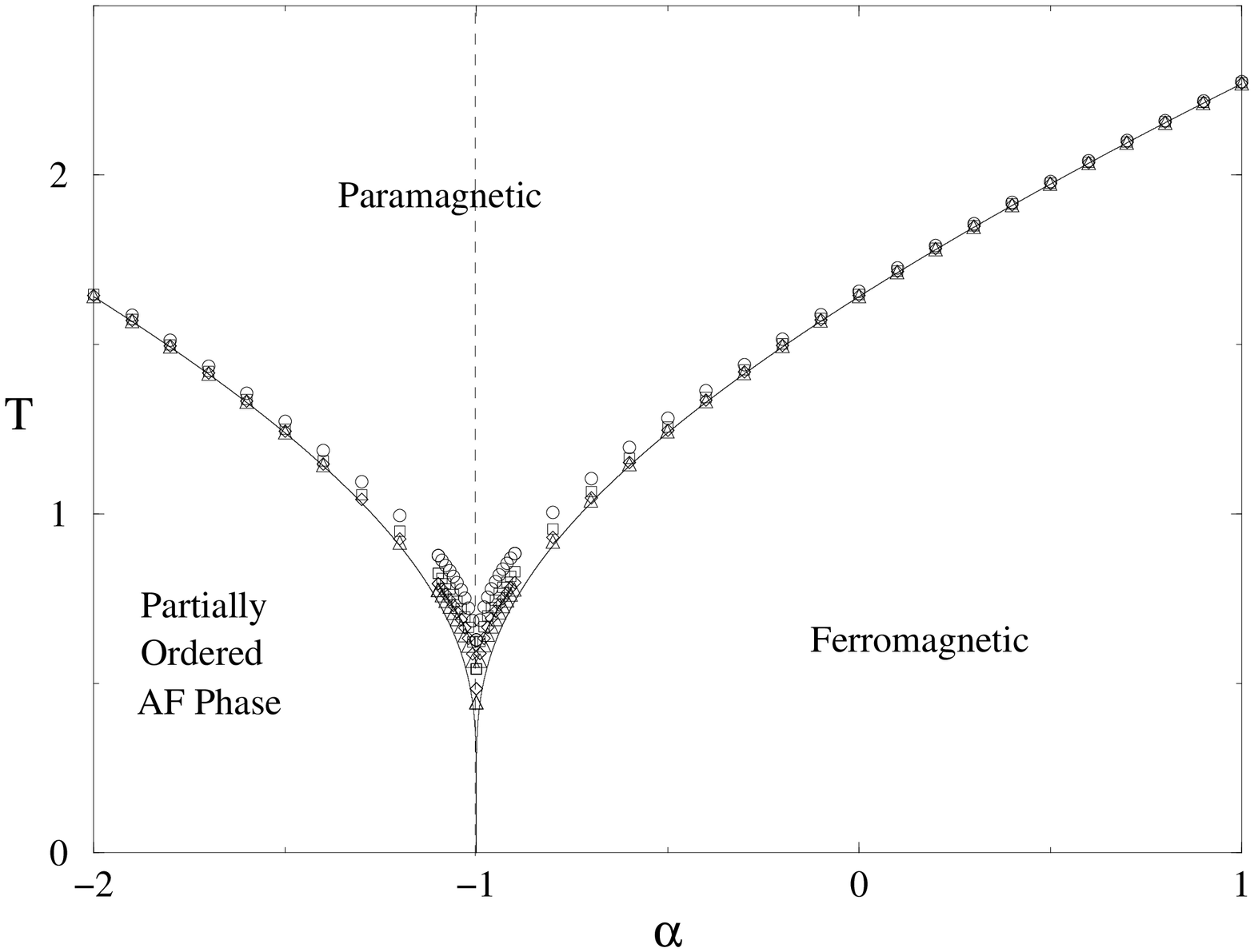}
\end{center}
{\large (B)}
\begin{center}
\includegraphics[width=10cm,angle=-90]{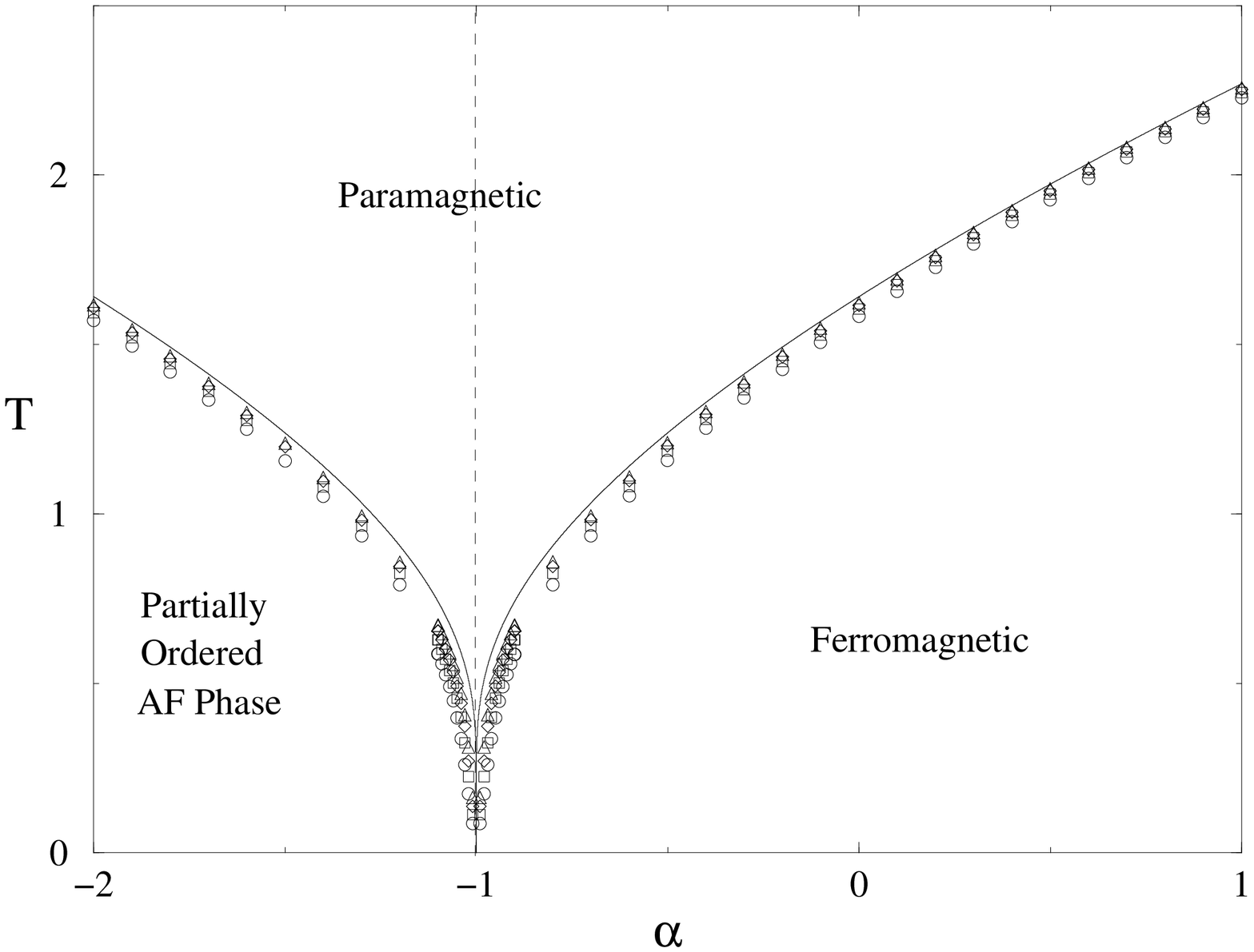}
\end{center}
\caption{Exact phase diagram for the Ising Piled-up-domino model, 
found by Andr\'e \etal\cite{abcc79} shown as a solid line along with
the finite size estimates calculated from the phenomenological
renormalisation group for widths 6 and 8 ($\opencircle$), 8 and 10
($\opensquare$), 10 and 12 ($\opendiamond$) and 12 and 14 ($\opentriangle$).
In (A) periodic boundary conditions are used, whilst in (B) open
boundary conditions are used.
The disorder line is shown as a dashed line and the phase boundaries
as solid lines.}
\label{isingpd}
\end{figure}

Before trying to interpret the phase diagrams shown in \fref{pudpd}
we review the main features of the Ising piled-up-domino model, reported
in~\cite{abcc79}. The
phase diagram, shown in \fref{isingpd}, 
is made up of two transition lines in the Ising
model universality class. The transition line  for $\alpha>-1$
corresponds to the standard Ising transition between a ferromagnetic
low-temperature phase and a paramagnetic high-temperature phase. For
$\alpha<-1$ the ground state is highly degenerate. The low temperature
phase is characterized as follows: the spins on the antiferromagnetic
lines order ferromagnetically in the $y$ direction 
and antiferromagnetically in the $x$ direction, where $x$ and $y$ directions are taken as in \fref{lattice}. The
spins on the horizontal ferromagnetic lines are partially disordered. 
The transition is between this partially ordered antiferromagnetic
phase and the high-temperature paramagnetic phase. The two transition
lines meet with infinite slope at $\alpha=-1$, where the critical
temperature is zero.

These results contrast with the phase diagrams shown in \fref{pudpd}
for the 
three-state Potts models. The phase diagrams shown are calculated
using finite width strips of width $L$ in the direction perpendicular
to the antiferromagnetic lines and infinite in the parallel
direction. In \fref{pudpd}~(A) periodic 
boundary conditions are taken
and in \fref{pudpd}~(B) free boundary conditions are taken. 
Periodic boundary conditions reduce the number of possible
configurations available to the system, compared to the free boundary
case, making it harder to excite the system, and tending to lock in
the low temperature phase. This is observed in the phase diagrams
where in the periodic boundary case the critical temperature estimates
converge to the thermodynamic limit from above, whilst for the open
boundary case the critical temperature estimates converge from below
(see also \tref{pudtc}). In \tref{pudtc} we compare the
critical-temperature estimates for the two boundary conditions for
various key values of $\alpha$ and we find good agreement between the
boundary conditions. 

The two estimated phase diagrams are, however,
quite different at first sight, notably for $\alpha$ less than about
$-1.3$. Before trying to reconcile the differences, let us summarise
the similarities. For $\alpha=-1$ there exists a
finite-temperature phase transition with $T_c=0.37\pm 0.01$. 
In \tref{pudnu} we show
finite-size estimates for the exponent $\nu$ calculated using the two
different boundary conditions. The extrapolated value of $\nu$ for the
open boundary conditions at $\alpha=-1$ indicates a transition in the
three-state Potts universality class. The finite-size estimates for
$\alpha=-1$ with periodic boundary conditions are still too far from
the thermodynamic limit to permit good extrapolation.

The point $\alpha=-1$ with $T=0$ corresponds to a
special point on the phase diagram, since it is the point which
separates the ferromagnetic ground state ($s=0$) from the frustrated
highly degenerate ground state ($s$ finite).
It therefore
corresponds to a transition point, which one would expect to be
connected to the rest of the phase diagram. 
The phase diagram with open boundary conditions clearly shows a
reentrant phase ending in a multiphase point which we identify with
the special point $T=0,\ \alpha=-1$.
 Whilst the periodic boundary results do show the begining of a
reentrant phase, the results are less clear. 
However, as we have seen, there is good agreement between the
extrapolated critical temperatures right up to $\alpha=-1$. The
transition lines become progressively more rounded as the system
width increases.  
The failure of the
transition lines to reach $T=0$ for the finite widths is probably
related to a stabilisation of the low-temperature phase by the
periodicity condition. To verify this conjecture, we compare the exact
phase diagram found by Andr\'e \etal\cite{abcc79} with both
periodic boundary condition and open boundary condition calculations, 
see \fref{isingpd}. 
The convergence is excelent in both cases but, as for the Potts model, 
the periodic boundary condition results fail reach $T=0$ 
(the transition lines for widths 12 and 14 reach $T=0.444$), 
whereas the open
boundary condition  transition lines do extend right down to $T=0$. 
In the light of these considerations it seems to us likely that the
point $T=0,\ \alpha=-1$ remains a multiphase point in the Potts model,
where the paramagnetic, ferromagnetic and partially ordered phases
meet. This view is conforted by the observation that the residual
entropy per site at zero temperature is higher for $\alpha=-1$ than
for either $\alpha<-1$ or $\alpha>-1$. This means that the ground
state is more disordered at $\alpha=-1$ than for $\alpha<-1$ presumably
due the the  coexistence of the paramagnetic and partially disordered 
phases.

In models with competing interactions, as is the case here, one may
define disorder lines which indicate 
where the nature of the correlation function
changes from monotonic (due to the ferromagnetic
interactions) to oscillatory (due to the antiferromagnetic
interactions)\cite{steph}. 
It is easy
to show that the correlation function (in the transfer direction) 
is related to the first and
second largest eigenvalues in  absolute value (the largest being
always positive) through:
\begin{equation}
G(x)\propto \left(\frac{\lambda_1}{\lambda_0}\right)^x.
\end{equation}
If $\lambda_1$ is positive, then $G(x)$ is monotonically decreasing,
but if $\lambda_1$ is negative then 
\[
G(x)=(-1)^x \left(\frac{|\lambda_1|}{\lambda_0}\right)^x,
\]
and is oscillatorily-decreasing. If we define $\lambda_1^+$ as the
largest positive eigenvalue, and $\lambda_1^-$ as the largest (in
absolute value) negative eigenvalue, then the disorder line is found
when $\lambda_1^+=|\lambda_1^-|$.

The disorder lines for various lattice widths are shown in
\fref{pudpd} as lines. It is common to find disorder lines associated
with reentrant phases. In this case the disorder lines must enter the
cusp caused by the reentrant paramagnetic phase;
since
the disorder line corresponds to a change in behaviour of the
correlation function it
may not cross a critical line. 
The disorder lines
for the two different boundary conditions may be seen to converge to a
common line as the thermodynamic limit is approached. 
The limiting disorder line is curved, contrary to the Ising case, see
\ref{diso},
were the disorder line is calculated analytically and
found to be a vertical line at $\alpha=-1$.  
The
two transition lines join in the form of a cusp, with the disorder
line between them (as in the Ising case). The form of the disorder
line may then be seen as confirming the existence of the reentrant
paramagnetic phase.

Phenomenological renormalisation indicates two possible phase diagrams
depending on which boundary conditions were taken: Either the
transition line extends to $\alpha\to -\infty$ and the phase diagram
resembles qualitatively  
a distorted version of the Ising model Phase diagram, or there is a
second reentrance. 
The true thermodynamic phase diagram
may correspond to one of the two posibilities discussed, or 
to a combination of the two, or indeed to neither. 

In order to shed some light on this issue
it is interesting to look at a
variety of average thermodynamic quantities, 
such as the magnetisation, energy, average numbers of each
type of bond that are satisfied, etc. 
It is possible to calculate with relative ease such local average
quantities in the framework of a transfer matrix calculation. 

In order to calculate the
magnetisation
it is
necessary to first calculate the probability of having configuration
${\cal C}_x$ in column $x$. This probability is simply the ratio of
the partition function restricted to having configuration ${\cal C}_x$
and the unrestricted partition function, which in terms of transfer
matrices may be written:
\begin{equation}
p({\cal C}_x)=\frac{
\Tr \left\{ T^x | {\cal C}_x \rangle\langle {\cal C}_x |
T^{N-x}\right\}
}{
\Tr \left\{T^N\right\}}.
\end{equation}
Writing $|{\cal C}_x \rangle$ in terms of the eigenvectors of $T$ and
taking $N$ to infinity and using the fact the $T$ is a real symmetric
matrix leads to:
\begin{equation}
p({\cal C}_x)=\langle 0 |{\cal C}_x \rangle^2.
\end{equation}
 The magnetisation for the Potts model is defined
as:
\begin{equation}
m = \frac{1}{q-1}\left(q \langle n_{\rm max} \rangle-1\right),
\end{equation}
with $n_{\rm max} $ the maximum of $\{\rho_i\}$, where
$\rho_i$ are the densities of spins in each of the Potts states. 
Defining $m({\cal C})$ as the magnetisation
for a column state as 
\begin{equation}
m({\cal C}) = \frac{1}{q-1}\left(q 
\frac{N_{\rm max}({\cal C})}{L}-1\right),
\end{equation}
with $N_{\rm max}$ the largest of the numbers of each spin type 
present in
 configuration ${\cal C}$. The magnetisation is then given as
\begin{eqnarray}\nonumber
m&=&\sum_{\cal C} m({\cal C}) p({\cal C})  \\
&=&\sum_{\cal C} m({\cal C}) \langle 0 | {\cal C} \rangle^2.
\end{eqnarray}

Since $\langle 0 | {\cal C} \rangle^2$ is the probability of finding a
column in configuration ${\cal C}$, then the probability of finding a
column with magnetisation $m$ is given by
\begin{equation} 
P(m)=\sum_{\cal C} \delta(m({\cal C})-m)\langle 0 | {\cal C}
\rangle^2.
\end{equation}

\begin{figure}
\begin{center}
\includegraphics[width=10cm,angle=-90]{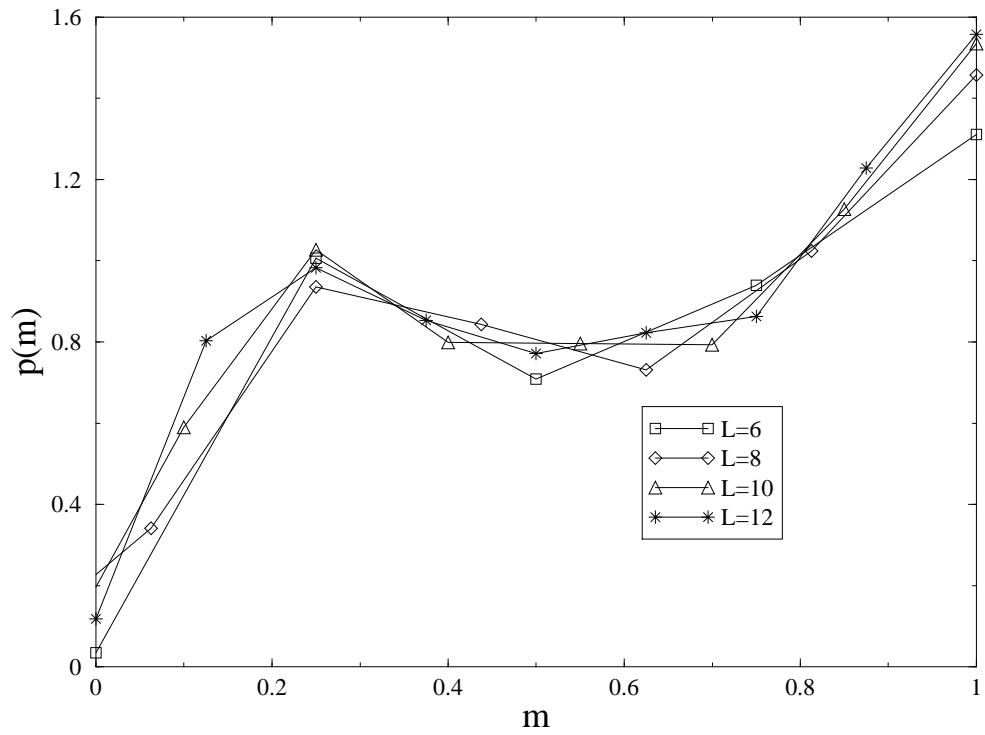}
\end{center}
\caption{Probability density for the magnetisation for $\alpha=-1.3$
and $T=0.2$}\label{prob}
\end{figure}

It is important to note that this is {\em not} the probability of
a magnetisation $m$ for the system as a whole. Since this probability
distribution is restricted to a column it gives a finer probe into the
probable column configurations. 
This is plotted for $\alpha=-1.3$ and
$T=0.2$ as a probability
density function in figure~\ref{prob} by dividing each probability by
the interval between allowed values of the magnetisation for a given
lattice width.  It may be clearly seen that the probability density is
developing two peaks, one at $m=1$ and a second at a substantially
lower value of $m$. This implies that in the low temperature phase
each column has a high probability of being ordered ferromagnetically
(along the $y$ direction)or of being (partially) disordered. How these
are arranged in the $x$ direction is not easy to calculate in the
transfer matrix formalism, but preliminary Monte Carlo
calculations\cite{dfgp} 
seem to indicate that the system alternates in the $x$ direction
between ferromagnetically ordered and disordered columns.
The probability distribution shown is
virtually identical for the two considered boundary conditions and for
$\alpha=-1.3$ and $\alpha=-2$. The peak at $m=1$ is lost in the high
temperature phase, clearly indicating the absence of magnetic
order. No evidence of two different frustrated phases
could be found from the distribution of the
magnetisation probability.

To calculate quantities such as the energy per spin or the densities
of satisfied bonds it is necessary to calculate the probabilities of
having a configuration ${\cal C}_x$ in column $x$ and ${\cal C}_{x+1}$
in column $x+1$ since these quantities are defined in terms of pairs
of spins, which may  belong to two adjacent columns. The calculation
of the joint probability $p({\cal C}_x,{\cal C}_{x+1})$ follows
exactly as above:
\begin{eqnarray}
p({\cal C}_x,{\cal C}_{x+1})&=&
\frac{
\Tr \left\{ T^x | {\cal C}_x \rangle\langle {\cal C}_x |
T | {\cal C}_{x+1} \rangle \langle {\cal C}_{x+1} |
T^{N-x-1}\right\}
}{
\Tr \left\{T^N\right\}}\\
&=& \frac{\langle 0 | {\cal C}_x \rangle\langle 0 | {\cal C}_{x+1}
\rangle \langle {\cal C}_x | T | {\cal C}_{x+1}\rangle}{\lambda_0} 
\end{eqnarray}
Defining $n_{\rm ferro}({\cal C}_x,{\cal C}_{x+1})$ and 
$n_{\rm antiferro}({\cal C}_x,{\cal C}_{x+1})$ as the densities of
satisfied bonds, we may calculate the average values:
\begin{equation}
\langle n_{\rm ferro} \rangle = 
\sum_{{\cal C}_x,{\cal C}_{x+1}} n_{\rm ferro}({\cal C}_x,{\cal
C}_{x+1})
p({\cal C}_x,{\cal C}_{x+1}).
\end{equation}
and similarly for $\langle n_{\rm antiferro} \rangle$.

\begin{figure}
\noindent{\large (A)}
\begin{center}
\includegraphics[width=10cm,angle=-90]{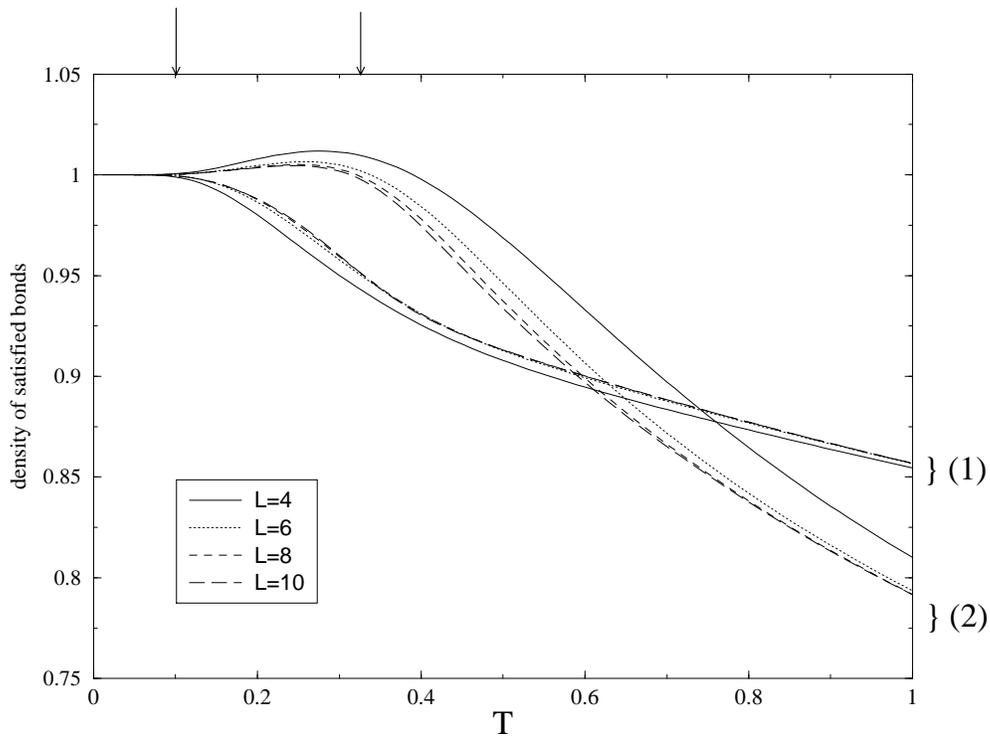}\\
\end{center}
\noindent{\large (B)}
\begin{center}
\includegraphics[width=10cm,angle=-90]{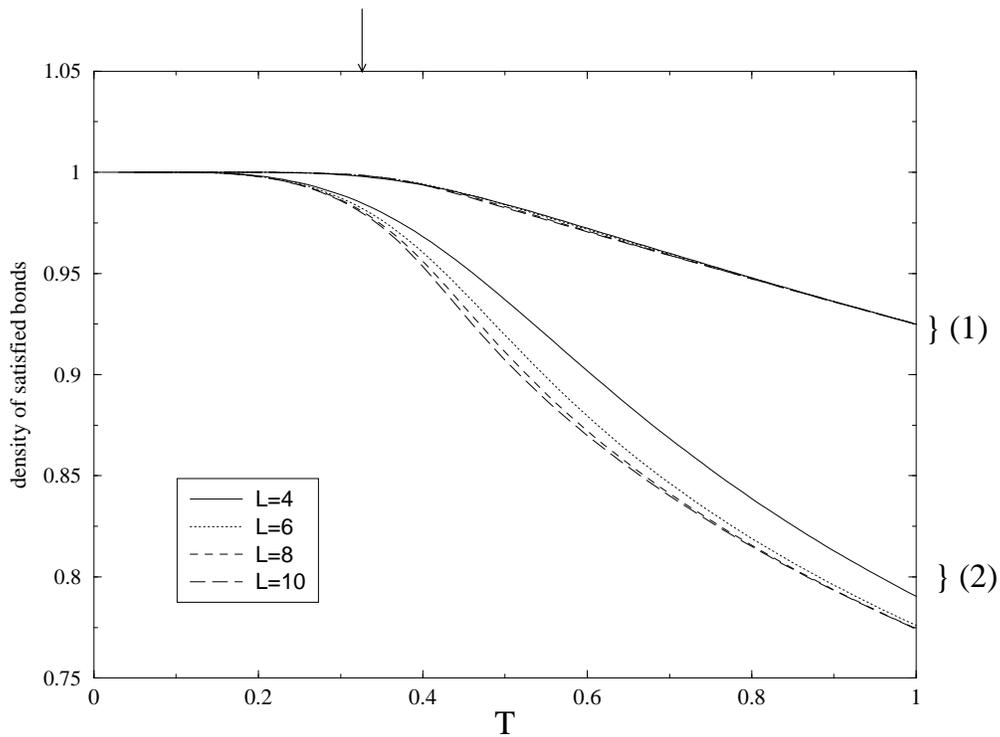}
\end{center}
\caption{The set of lines in each graph labeled (1) shows 
the number of
satisfied antiferromagnetic bonds normalised by the total number of
antiferromagnetic bonds, while the set of lines labeled (2) 
shows the number of
satisfied ferromagnetic bonds per spin. Figure (A) corresponds to
$\alpha=-1.3$ and figure (B) to $\alpha=-2$ both with periodic
boundary conditions.}
\label{dens}
\end{figure}

These two quantities are plotted in \fref{dens} for (A) $\alpha=-1.3$
and (B) $\alpha=-2$ with periodic boundary conditions. 
It may be seen that
there is a clear difference of behaviour between $\alpha=-1.3$ 
and $\alpha=-2$. In the first case the two densities may be seen to drop
away from their maximal values at the different temperatures
(shown approximately by
arrows in \fref{dens} (A)). 
In the
second case this same drop off occurs at the same temperature for the
two different densities. 
The results
for the densities show a high
degree of agreement between the two different boundary
conditions. Both show a qualitative difference of behaviour between
$\alpha=-1.3$ and $\alpha=-2$, and evidence of possible phase
transitions for both values of $\alpha$, indicated by arrows in
figure~\ref{dens}, two for $\alpha=-1.3$ and one for $\alpha=-2$.
The specific heat does not diverge with system size for either of
these values
of $\alpha$, but the peaks of the specific heat converge to a 
temperature consistent with the critical temperature shown in
\fref{pudpd} (A) for $\alpha$ less than about $-1.5$ and all $\alpha\ge -1$.

\begin{figure}
\begin{center}
\includegraphics[width=10cm,angle=-90]{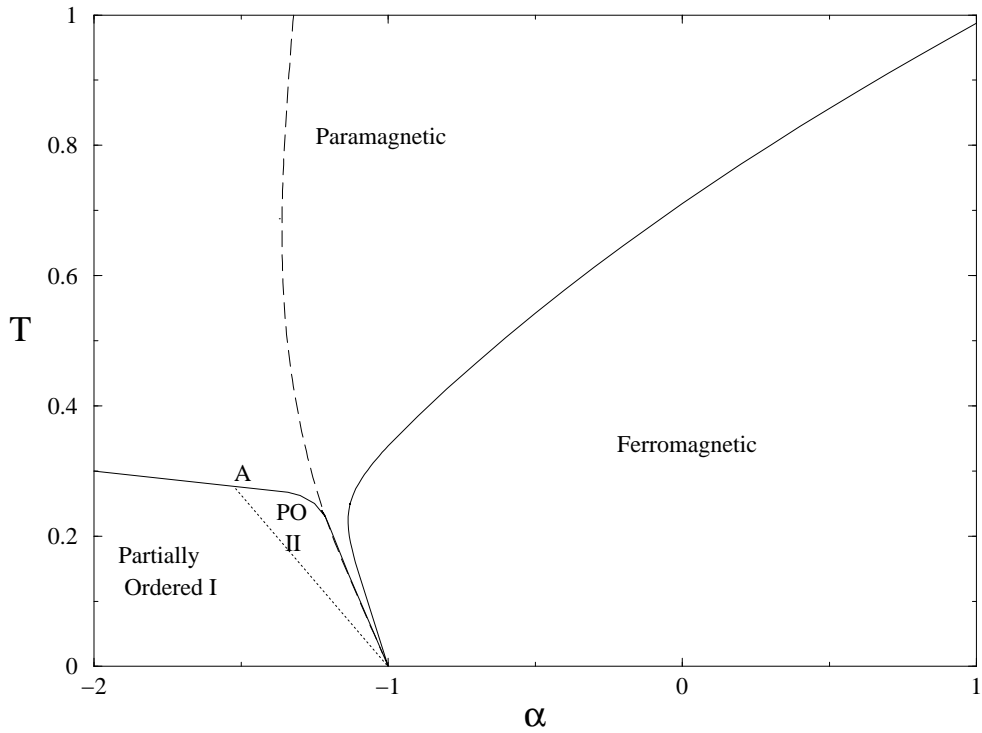}
\end{center}
\caption{Possible thermodynamic phase diagram for the three-state
Potts piled-up-domino  model.
The disorder line is shown as a dashed line, and the dotted line is a
possible extra transition line between two partially ordered phases
(see text) with a possible multicritical point, marked as A on the
phase diagram.}
\label{conpd}
\end{figure}

We conclude that there is sufficient evidence to conjecture a phase
diagram as shown
semi-schematically in \fref{conpd}, with a possible second partially ordered
phase at low temperature, (PO II). This additional phase, compared
to the piled-up-domino Ising model, is only presented speculatively; 
the width of this phase 
reduces as system size increases (see \fref{pudpd}) and may disappear
in the thermodynamic limit. If such a phase were to exist then it is
likely to terminate at a multicritical point (A). The presence of such
a multicritical point could provide evidence of such a phase.

\section{Conclusions}\label{conc}

The main results presented in this article relate to the
Potts piled-up-domino model, where we have presented behaviour which
is remarkably different from the equivalent Ising case, notably a
finite-temperature transition and reentrant behaviour. It is unusual
to see reentrance in such a simple model. Models showing reentrance
normally contain competing nearest-neighbour and
next-nearest-neighbour 
interactions\cite{dpbook} 
or interaction
between different spin types~\cite{potts_re}. The additional entropy
available due to the greater number of spin states stabilises the
ferromagnetic ground-state as the temperature is increased (with
$\alpha=-1$), 
contrary to the na\"{\i}ve anticipation that
the extra degrees of freedom will disorder the system more. This may
be understood as follows: in the Ising case the ground states may be
mapped one to one onto dimer coverings of the lattice, but when the
$q$-state Potts model is considered the mapping is no longer one to
one, and the additional degeneracy is not uniformly
distributed. This reweighting must apply to all the low lying states,
and must serve to stabilise the ferromagnetic phase as the temperature
is increased through an order by disorder mechanism. 
In the zig-zag case the different dimer coverings have
the same weight in the ground state, as in the Ising case, and we do
not observe an equivalent stabilisation of the ferromagnetic ground
state.

The exact nature of the phase diagram and the low temperature phases
for $\alpha<-1$ remain to be fully determined, and will doubtless
require other methods. This will be the object of a future study.

\ack
We would like to thank H.T. Diep, C. Pinettes  and T.T. Truong 
for interesting and useful 
discussions.

\appendix

\section{Disorder line for the Ising piled-up-domino model}\label{diso}

The free energy and phase diagram for the 
Ising piled-up-domino model was solved exactly by Andr\'e
\etal\cite{abcc79}, but to the best of our knowledge the disorder line
has not been found for this model.

Andr\'e \etal give the (dimensionless) free energy per spin as:
\begin{equation}
f=\log 2 +\frac{1}{4\pi^2} \int_0^\pi\int_0^\pi {\rm d}q_x{\rm d}q_y
\log \Delta(q_x,q_y),
\end{equation}
with 
\begin{eqnarray}\nonumber
\Delta(q_x,q_y)&=&\frac{1}{2}[\cosh 2\beta J \cosh 2\alpha\beta J
+\cosh^2 2\beta J\cosh 2(1+\alpha)\beta J\\
& &-\sinh^2 2\beta
J\cos q_y
-2\cosh 2\beta J\sinh 2(1+\alpha)\beta J\cos q_x\\\nonumber
& &+\sinh 2\beta
J\sinh 2\beta \alpha J\cos 2q_x],
\end{eqnarray}
from which we deduce the form of the generic correlation function as 
\begin{equation}
G(q_x,q_y)=\frac{1}{\Delta(q_x,q_y)},
\end{equation}
leading to
\begin{equation}
G(x,y)=\frac{1}{4\pi^2}\int_0^\pi \int_0^\pi {\rm d}q_x{\rm d}q_y
\frac{\exp i(xq_x+yq_y)}{\Delta(q_x,q_y)}.
\end{equation}

The modulation of $G(x,y)$ in the $x$ ($y$) 
direction is governed by the maximum of
$G(q_x,q_y)$, or the minimum of $\Delta(q_x,q_y)$, 
with respect to qx (qy). 
Since the modulation
is anticipated in the $x$-direction (see\cite{abcc79}), we seek the
turning points of $\Delta(q_x,q_y)$ with respect to $q_x$:
\begin{equation}
\frac{\partial \Delta}{\partial q_x}=2\cosh 2\beta J\sinh
2(1+\alpha)\beta J \sin q_x - 2\sinh 2\beta J \sinh 2\alpha \beta J
\sin 2 q_x,
\end{equation}
which becomes zero when $q_x=0,\ \pi$ or when
\begin{equation}\label{third}
\cos q_x=\frac{\cosh 2\beta J \sinh 2(1+\alpha)\beta J}{2 \sinh 2
\beta J \sinh 2 \alpha\beta J}.
\end{equation}

Checking the second derivative of $\Delta$ it may be seen that the
$q_x=0$ solution is a minimum as long as 
\begin{equation}
\cosh 2\beta J\sinh 2(1+\alpha)\beta J > 2\sinh 2\beta J\sinh
2\alpha\beta J,
\end{equation}
and the $q_x=\pi$ solution is a minimum as long as 
\begin{equation}
\cosh 2\beta J\sinh 2(1+\alpha)\beta J > -2\sinh 2\beta J\sinh
2\alpha\beta J.
\end{equation}
These two conditions correspond exactly to the boundaries within which 
the solution
\eref{third} exists, which must correspond to a maximum.

In the range of values of $\alpha$ for which the
two solutions $q_x=0$ and $q_x=\pi$ are minima it remains to determine
which corresponds to
the absolute minimum. This may be determined by considering 
\begin{equation}
\Delta(\pi,q_y)-\Delta(0,q_y)=4\cosh 2\beta J\sinh 2(1+\alpha) \beta J.
\end{equation}
This changes sign when $\alpha=-1$. For $\alpha>-1$ the leading mode
of the correlation
function is the $q_x=0$ mode, while for $\alpha<-1$ the leading mode
is the $q_x=\pi$ mode. This gives the
location of the  disorder line as $\alpha=-1$ for
all temperatures.

\Figures

\Tables

\begin{table}
\caption{Ground-state entropy for the Potts piled-up-domino
model}\label{pudent} 
\begin{indented}
\item[]\begin{tabular}{@{}lllll}
\br
$L$ & $\alpha<-1$ & $\alpha<-1$ & $\alpha=-1$  & $\alpha=-1$ \\
 & Free b.c.'s & Periodic b.c.'s & Free b.c.'s & Periodic b.c.'s \\
\mr
2 & 0.5025263 & 0.5025263 & 0.635098 & 0.6350983 \\
4 & 0.3854409 & 0.3588588 &0.471894 & 0.4333004 \\
6 &0.3521346& 0.3238563 & 0.430075 &  0.3934622 \\
8 & 0.3369267 &  0.3113484 & 0.412780 &  0.3825248 \\
10 & 0.328266 & 0.3057865 & 0.403783 & 0.3789744 \\
12 &  0.3226777 &  0.3028948 &  0.398421 &   0.3776883 \\
\mr
$\infty$  & 0.2999 & 0.2972  & 0.3814 & 0.3766 \\
\mr
Ising & $\frac{1}{2}\log((1+\surd 5)/2)$ &  & $\frac{G}{\pi}$ &   \\
Exact  & $=0.240659\cdots$ &  & $0.2916\cdots$ &   \\
\br
\end{tabular}
\end{indented}
\end{table}

\begin{table}
\caption{Critical temperature estimates for the zig-zag model.}\label{zigtc} 
\begin{indented}

\item[]\begin{tabular}{@{}lll}
\br
$L/L^{\prime}$ & $\alpha=1$ &\ \ \ $\alpha=0$\ \ \  \\
\mr
4/6 &\ 1.0028517\ &\ 0.679435201\ \\
6/8 &\ 0.997615939\ &\ 0.675820268\ \\
8/10 &\ 0.996228353\ &\ 0.674737229\ \\
10/12 &\ 0.99569457\ &\ 0.674343044\ \\
\mr
$\infty$ &\ 0.995\ &\ 0.674\ \\
\mr
Exact &\ $1/\log(1+\surd 3)$\cite{rgb70}\ & (1) \\
&\ $=.994973\cdots$\ & $=.6737596022\cdots$ \\
\br
\end{tabular}
\item[] (1) Critical temperature for $\alpha=0$ solution of
$e^{3/T}-3e^{2/T}-6e^{1/T}-1=0$.
\end{indented}
\end{table}

\begin{table}
\caption{ Critical temperature estimates for the piled-up-domino
model}\label{pudtc}
\begin{indented}
\item[]\begin{tabular}{@{}llllllll}
\br
 & \centre{2}{$\alpha=0$} &  \centre{3}{$\alpha=-1$} & 
\centre{2}{$\alpha=-1.3$} \\\ns
$L/L^{\prime}$ & \crule{2} &\crule{3} &\crule{2} \\\ns
 & PBC & FBC & PBC & PBC-y 
& FBC & PBC & FBC \\
\mr
2/4 &  0.764578 & 0.623342 & 0.482104 & 0.428436 & --- & --- & 0.081726 \\
4/6 &  0.739625 & 0.677582 & 0.456832 & 0.416958 & 0.272384 & 0.350795
& 0.189215 \\  
6/8 &  0.727985 & 0.696787 & 0.417748 & 0.393249 & 0.310488 & 0.329505
& 0.235781 \\ 
8/10 & 0.724292 & 0.705696 & 0.395310 & 0.381282 & 0.328203 & 0.315065
&0.253698 \\ 
10/12  & 0.722934 & 0.710524 & 0.383267 & 0.367335  & 0.338479 &
0.305799 & 0.262133 \\ 
\mr
$\infty$   & 0.722 & 0.722 & 0.361 &  0.367 & 0.379 & 0.283 & 0.274 \\ 
\mr
exact & \centre{2}{$1/\log(4)=.721358\cdots$} & --- & --- & --- & --- &  --- \\
\br
\end{tabular}
\item[]PBC: Periodic boundary conditions
\item[]FBC: Free boundary conditions.
\item[]PBC-y: Periodic boundary conditions with transfer direction
perpendicular 
to dashed lines (see Figure~\ref{lattice})
\end{indented}
\end{table}

\begin{table}
\caption{Estimates for the exponent $\nu$ for the piled-up-domino
model}\label{pudnu}
\begin{indented}
\item[]\begin{tabular}{@{}lllllll}
\br
& \centre{3}{Periodic Boundary Conditions} &\centre{3}{Free Boundary Conditions} \\\ns

$L/L^{\prime}$ & \crule{3} & \crule{3}\\
& $\alpha=1$ & $\alpha=0$ & $\alpha=-1$ & $\alpha=1$ & $\alpha=0$ & $\alpha=-1$ \\
\mr
2/4 &  0.807155 & 0.727115 & 0.457417 &  0.897734 & 0.770501 & --- \\
4/6 & 0.827697 & 0.754573  &0.505535 & 0.891758 & 0.836442 & 2.049413\\
6/8 & 0.837053  & 0.792602 & 0.573436 & 0.883714  & 0.852665 & 1.354167\\
8/10 & 0.839283 & 0.811782 & 0.639087 & 0.877283 & 0.857856 &  1.162991\\
10/12 &  0.840722 & 0.821423  & 0.697795 &  0.872277 & 0.859395 &  1.075589\\
\mr
$\infty$ & 0.842 & 0.837 & 1.139 & 0.836 & 0.860 & 0.872 \\
\br
\end{tabular}
\end{indented}
\end{table}

\end{document}